\documentstyle[aps,epsfig,multicol]{revtex}
\begin{document}
\draft
\title {Small-world phenomena and the statistics of 
linear polymers }
\author{Parongama Sen$^1$ and Bikas K. Chakrabarti$^2$ }

\address{$^1$Department of Physics, University of Calcutta,
92 A.P. C. Road, Calcutta 700009, India.\\
e-mail paro@cucc.ernet.in \\
$^2$ Saha Institute of Nuclear Physics, 1/AF Bidhan Nagar, Calcutta 700064,
India.\\
e-mail: bikas@cmp.saha.ernet.in}
\maketitle
\begin{abstract}
 A regular lattice in which the sites can have long range  
connections at a distance 
$l$ with a probabilty  $P(l) \sim l^{-\delta}$, in addition to the short
range nearest neighbour connections, shows small-world behaviour for 
 $0 \le \delta < \delta_c $. 
In the most appropriate physical example of such a system, 
namely the  linear polymer network, the exponent $\delta$ is
related to the exponents of the corresponding $n$-vector model
in the $n \rightarrow 0$ limit, and its value is less than $\delta_c$.
Still, the polymer networks do not show small-world behaviour.
Here, we show that this is due a (small value) constraint on the
number $q$ of long range connections per monomer in the network. 
In the general $\delta - q$  space,  
we obtain a phase boundary separating regions with and  without
small-world behaviour, and show that the polymer network falls
marginally in the regular lattice region.
\end{abstract}


\begin{multicols}{2}

A small-world network (SWN) develops out of regular lattices
having local connections with additional long range links or connections 
with a finite probability. It acquires the intriguing property  that while
the local connectivity structure remains similar to the original 
underlying lattice, the shortest path between any two
lattice sites becomes similar to that of random graphs \cite{Watts,newman}.
Specifically, if we take a regular $d$-dimensional lattice, then the 
neighbourhood of any site of the lattice is given by the structure of the 
lattice
and the shortest path between any two sites of the lattice  is
of the order of $N^{1/d}$ where $N$ is the total number of sites in the
lattice. For a 
random graph, having random connections at all ranges, a similar shortest
path distance between any two points grows as $\ln N$ 
implying that the effective lattice dimensionality
$d \rightarrow \infty$. The local structure of a random graph is 
completely amorphous. Watts and Strogatz \cite{Watts}
in their model showed that starting from a regular lattice, when some of
the nearest neighbour bonds are replaced by long range connections at
random with a finite probability $p$,  the network shows an interesting
feature: while the local connectivity remains practically the same as 
in the regular
lattice, the global shortest path distance $S_N$ scales as $\ln N$ for
a network of size $N$ for any non-vanishing $p$.

This property of the SWN is of importance in various social communication
networks like internet links, disease spreading etc. This also explains the 
amazingly low value of the number of  steps connecting any two  members of 
such networks, like the
`six degrees of seperation' observed in Milgram's experiment and similar
situations \cite{Milgram}. The minimal step number connecting two members 
of these networks scales as $\ln N$, rather than as $N$, the total
population of the network, while the local coordination
structure remains almost the same. Recently, the propagation of 
thermal correlations 
in such networks have also been studied; particularly, the effect of 
such random long range connections on the nearest neighbour interacting 
Ising chain \cite{Barrat,Gitterman}. 
With long range interactions occurring  with probability $p$, the transition 
temperature becomes non-zero for $p> p_c$ for the Ising chain; $p_c$ = 0
according to Barrat and Weigt \cite{Barrat} and $p_c \neq 0$  according to
Gitterman \cite{Gitterman}. 
Estimates of
this transition temperature and the nature of the correlations near the
transition point have been investigated in the above-mentioned
studies \cite {Barrat,Gitterman}. 

In such SWN, the probability $p$ of
the long range connections is simply the fraction of such connections
which are added randomly and are independent of the 
range. Recently, in some extended SWN model, the probability 
$P(l)$ that two points at a distance $l$ along the chain are connected 
has been  taken to be 
dependent on $l$ such that  $P(l)  \sim l^{-\delta}$; $\delta=0$ corresponds to the
original model \cite{Watts}. These studies indicate that
for $\delta \geq  \delta_c$ the network effectively 
reduces to the regular lattice, while for $\delta \leq \delta_c$
it effectively becomes a SWN. There is apparently some disgreement
over the value of $\delta_c$: according to \cite{blumen} $\delta_c =2 $,
while in \cite{kleinberg} $\delta_c$ is found to be equal to $d$. 
Our  study here is essentially for one dimensional chains and we find
that for the corresponding unrestricted cases (defined later), 
 $\delta_c$ is equal to 2.

It appears that the most appropriate example of SWN in physics
is the example  of linear
polymers or the self-avoiding walk (SAW) model \cite{degennes}.
Here, the long range connections over the nearest neighbour
monomer-monomer connections (called the ``streets")
come from the random folding (forming the local 
loops or the ``bridges") 
of the chain in the embedding dimension $d$.
Fig. 1 shows how long range connections develop out of random
folding in a SAW. 
In fact, the structure of such a network has been studied intensively
some time back \cite{bkc1,bkc3,bkc5}.
Here of course, the probability $P$ of a connection at a (long) range $l$
 is given by the SAW statistics: $P(l) = G^{SAL}_l/G^{SAW}_l$
where 
$G^{SAW}_l = \mu^l l^{\gamma -1}$
is the number of SAW's 
of length $l$ and 
$G^{SAL}_l = \mu^l l^{-2+\alpha}$
 is the number of 
loops of length $l$ \cite{degennes}. Here $\mu$ is the connectivity 
coefficient of the SAW 
on the embedding lattice  and $\gamma$ and $\alpha$ are respectively the
susceptibility and specific heat exponent of the equivalent 
$n-$vector model in the limit $n \rightarrow 0$.
Hence 
\begin{equation}
P(l) \sim l^{-\delta},
\end{equation} 
where $\delta = 1+\gamma -\alpha $.
Using the approximate Flory formula $\nu = 3/(2+d)$ for the correlation length 
exponent and $\gamma \simeq 1$, we get $\delta \simeq 
 d\nu = 3d/(2+d)$. Hence $\delta < 2$ in $ d = 2$,
and is equal to 2 for $d = 4$.
In fact, with the best estimates of $\alpha$ and  $\gamma$ 
\cite {degennes}, one would also get
$\delta <2$ for $d \leq 3$ and $\delta =2$ for $ d = 4$.  The average
number $q$ of such additional long-range connections per site can also
be estimated easily \cite {degennes} : $q = (z -1) - \mu$, where $z$ is the 
lattice coordination number. This is because, out of the $z - 1$ options
left for the SAW to grow for the next step, on an average $\mu$ options
are chosen; the rest being visited earlier due to random folding of the
chain. With $\mu \simeq$ 2.638, 4.151 and 4.686, one gets $q \simeq$
0.36, 0.85 and 0.31 in square, triangular and simple cubic lattices
respectively.

The above comparison and the observations \cite {bkc3,bkc5}
on the SAW networks indicate that the 
requirement $P(l) \sim l^{-\delta} $ with $\delta < 2$ in $d <4 
$ for small-world effects, is not sufficient, as  
the small-world phenomena is certainly not observed in such SAW
networks. In particular, it was found that the shortest path
length $S_N$ in a SAW network of $N$-steps grows only linearly
in $N$ for any finite range interactions or bridges \cite {bkc5}. 
The SAW network therefore always remains a linear one, and no small-world
effect ($S_N \sim \log N$) or, for that matter, no extra dimensional effect
($S_N \sim N^x, x < 1$) can be seen.
Of course
the structures of the linear polymer network and the small-world 
network are  inherently  different,  although the effective probability 
of connections in both  cases are given by the same power law (1). In 
SAW network, there
is a structural restriction in the total number of neighbours: at any
point along the SAW chain, the total number of connections $q+2$ 
(as there are $q$ long range connections and two short
range connections for each site)  
 cannot exceed the coordination number of the underlying lattice. 
In SWN, however,  no such
restriction exists (theoretically $q$ here can go upto $N-3$ at each site
for $\delta =0$). 
In the SAW network, there exists also a correlation between the 
bridges. Such correlations are absent in SWN (see e.g., in 
Fig. 1, two long connections lie very close: such configurations are more
likely in a SAW than in a general SWN). 

Here we explore the differences that emerge from the constraint on  the
total number of long range connections  in SAWs to find whether
it is responsible for the non-SWN like behaviour of the SAW networks.
Precisely, we investigate the crossover from small-world 
 behaviour  to regular lattice behaviour indicated by the 
 crossover  in the shortest path
$S_N \sim$ from $\log N$  to   $N^{(1/d)}$,
in the parameter space of $\delta$ and $q$.

For a fixed finite average number $q$ of long range 
connections (per site) in the SWN, we vary
$\delta$ to obtain $\delta_c(q)$ at which the $S_N$ behaviour changes. We
then vary $q$ in the range $0 < q < 2$, which corresponds to the real
physical situation of SAWs in a square lattice.
Note that for the infinite chain, the probability
in (1) is easily normalisable for any $\delta$ greater than unity. 

In the SAW, the values of $\delta$ and $q$ are independent.
While treating both $\delta$ and $q$ as independent quantities in the
model where the probability of a long range connection of length $l$
varies as $l^{-\delta}$, it should
be mentioned that strictly speaking the total number of long range 
connections $qN$ 
depends on the value of $\delta$. For example, for $q \sim N$, one does 
not have a choice for the value of  $\delta $ as large values of $\delta$
will not be allowed in this case.
 Therefore, when  $q$ is  taken to be independent 
of $\delta$, the range of both parameters get restricted.
We find that making $ 0 <q  < 2$ is safe for values of $\delta  \leq 2.5$
in the sense that the desired scaling of the bond distributions is
intact.

We first generate a linear chain of length $N $. We then put additional
long range interactions (bridges) following the probability distribution
(1), with the restriction that the total number of such connections
 is $qN$. 
We then use a greedy algorithm  (see e.g., \cite{kirk}) to find out the 
shortest path
through the streets and bridges and count the number of steps $S_N$ 
connecting the end points of the SAW network.

In the unrestricted case (large $q$),
we find that the (phase) transition  from log to  
linear scaling is recovered 
at $\delta \geq 2$. This agrees with the result of \cite{blumen}. 
For the finite $N$ values considered here 
($N \le 10000$), at low  values of $\delta ~(< 1.4)$,
 the logarithmic scaling is clearly observed. 
However, for the intermediate values of $\delta ~(1.4 < \delta < 2)$, 
the variation 
is apparently power law like. We believe that this  is only
an effective behaviour and the
logarithmic behaviour will follow
 as $N \rightarrow \infty$. 
The justification is a-posteriori: from the results
for the constrained cases to be discussed later.

The constraint  that the total number of long range bonds for the
whole chain is $qN$ is a global constraint. It may also
be demanded that these bonds are distributed such that each site is not 
allowed to have more than
a fixed number of long range bonds; this is a local constraint.  
In a SAW embedded in a square lattice, 
there is a  local constraint as individual sites cannot have more than two 
long range neighbours.
We have studied both the locally and globally conserved cases; in the 
former the number of maximum long range bond at every site is
kept equal to two in analogy with the SAW (on a sqaure lattice) and keeping
the total number of long range connection equal to $qN$ at the same time.
However, the results with local and non-local 
conservation  are observed to be the same,
 as far as the
scaling  of the shortest path is concerned. 
The total path length in the locally restricted case
is smaller because  connections which are  redundant
in the globally restricted case are to some extent lesser
in number.

The variation of the shortest path with the path length
for a given $q$ and various $\delta$ values shows that 
when extrapolated for large $N$, there are only two kinds of behaviour: logarithmic and 
linear. The linear behaviour is obtained for $\delta > \delta_c(q)$. 
The logarithmic behaviour is quite apparent for small $\delta$.
 For intermediate 
values of  $ \delta < \delta _c(q)$, apparently 
there is an effective power law behaviour $S_N \sim N^x$; but we find that
with  increasing $N$, $x$ decreases and 
 one finally obtains the logarithmic behaviour. 
On the other hand, for $\delta \geq \delta_c$, $x$ is observed to remain practically
near unity.
This observation may require really large values of 
$N$ (e.g., $N > 10000$ for $q < 0.1$).
The transition boundary is difficult to
estimate very accurately; a problem encountered in many small-world
networks \cite{blumen}. 
At small values of $q$, we find that the shortest path becomes linear
at a value of $\delta_c(q) < 2$. 
As $q$ is increased, $\delta_c(q)$ increases,
and finally $\delta_c (q) $ approaches a value  2.0  as $q$ becomes large
(see Fig. 3).

In summary, our generalised model in which  both $\delta$ and $q$ are treated 
independently shows that the transition  from small-world  to the 
regular lattice  behaviour  occur across a
phase boundary  $\delta_c (q)$.
Extensive earlier investigations  on SAW networks had
established \cite{bkc3,bkc5} that they do not show any small-world
behaviour. In view of the recent studies \cite {blumen,kleinberg} on models 
having long range connections with probability distribution (1), and the
observation that they show small-world behaviour for $\delta < 2$, together
with the fact that $\delta$ is indeed less than two for SAWs, an apparent
contradiction arises. We resolve this here  by studying 
the phase diagram of a generalised model having long range connection 
probability distribution (1) and a restricted average number $q$ (per
site) of such additional long range connections. We show that a crossover 
from small-world to regular lattice behaviour occurs as one crosses
the phase boundary given by $\delta_c(q)$ ($\delta_c \rightarrow 2$ for
large $q$).  The points ($q, \delta$) corresponding to the square,
triangular and cubic lattices in Fig. 3 show that linear polymers
on these lattices lie marginally in the regular lattice region.
 We claim that the   
smallness of the  value  of $q$ ($< 1$)
for linear polymers makes it fall in the regular lattice region in the 
phase diagram,
and excludes the  possibility of its small-world behaviour. 
 As mentioned earlier, we have not taken into account the various
correlations developing in the SAW and thus the $q$ values 
in SAW and the network studied in the present paper may not be quantitatively
identical.  Secondly, the values of $q,\delta$ shown in Fig. 3
for the different lattices are theoretical estimates 
only and therefore  these points may actually lie deeper in the regular lattice
region.  Hence  we believe that the important result that
$\delta_c(q) < 2$ for small values of $q$ is responsible for 
the non-SWN like behaviour in SAW.

Acknowledgements: We thank A. Blumen and S. Jespersen
for their correspondences and comments on the manuscript. We are
grateful to S. M. Bhattacharjee for a critical reading
of the manuscript and comments. PS acknowledges DST (India) grant no. SP/S2/M-11/99.

\narrowtext

\begin{figure}
\caption{(a) Portion of a 
SAW network: the solid line represents the polymer or the street
and the dashed lines represent the nearest neighbour bridges.
(b) A stretched chain equivalent: contains long range connections 
with probability $P(l)$.}
\end{figure}
\begin{figure}
\caption{The behaviour of the shortest path length $S_N$ over $N$ steps 
along the chain. The curves are drawn for higher to lower values
of $\delta$ from top to bottom: (a)  $q= 0.01$; $\delta = 1.6,1.5,1.4,
1.2$ and $1.0$. The curves become linear for $\delta = 1.5 $ and above.
(b) $q = 0.02$; $\delta = 1.7,1.6,1.5,1.4,1.2$ and $1.0$. The
curves are linear above $\delta = 1.6$.
(c) $q = 0.2$; $\delta = 2.0,1.95,1.9,1.8,1.7$ and $1.6$. Linear behaviour is
observed above $\delta=1.90$. (d) $q= 0.5$; $\delta = 1.8,
1.9$ ,1.95,  2.0 and 2.1.
Linearity appears   above   $\delta =  2.0$.  
In (c) and (d)  where larger chains ($N=10000$) have been
considered, the change from the apparent power law to the actual logarithmic
behaviour below $\delta_c$ is clearly observed. The logarithmic
behaviour at $\delta <\delta_c$ can be observed at smaller values of $N$ 
as $q$ increases.}
\end{figure}
\begin{figure}
\caption{The regions corresponding to the different behaviour of
$S_N$ in the $\delta - q$ plane. The dashed line is a guide to the eye
 separating the small-world  region from the linear region.
 At low values of $q$, there is
 no small-world 
like phenomena even though $\delta $ is less than 2.0. 
The values of $q$ and $\delta$  for linear polymers 
on the square, triangular and cubic
lattices are shown by different symbols.  }
\end{figure}

\end{multicols}

\psfig{file=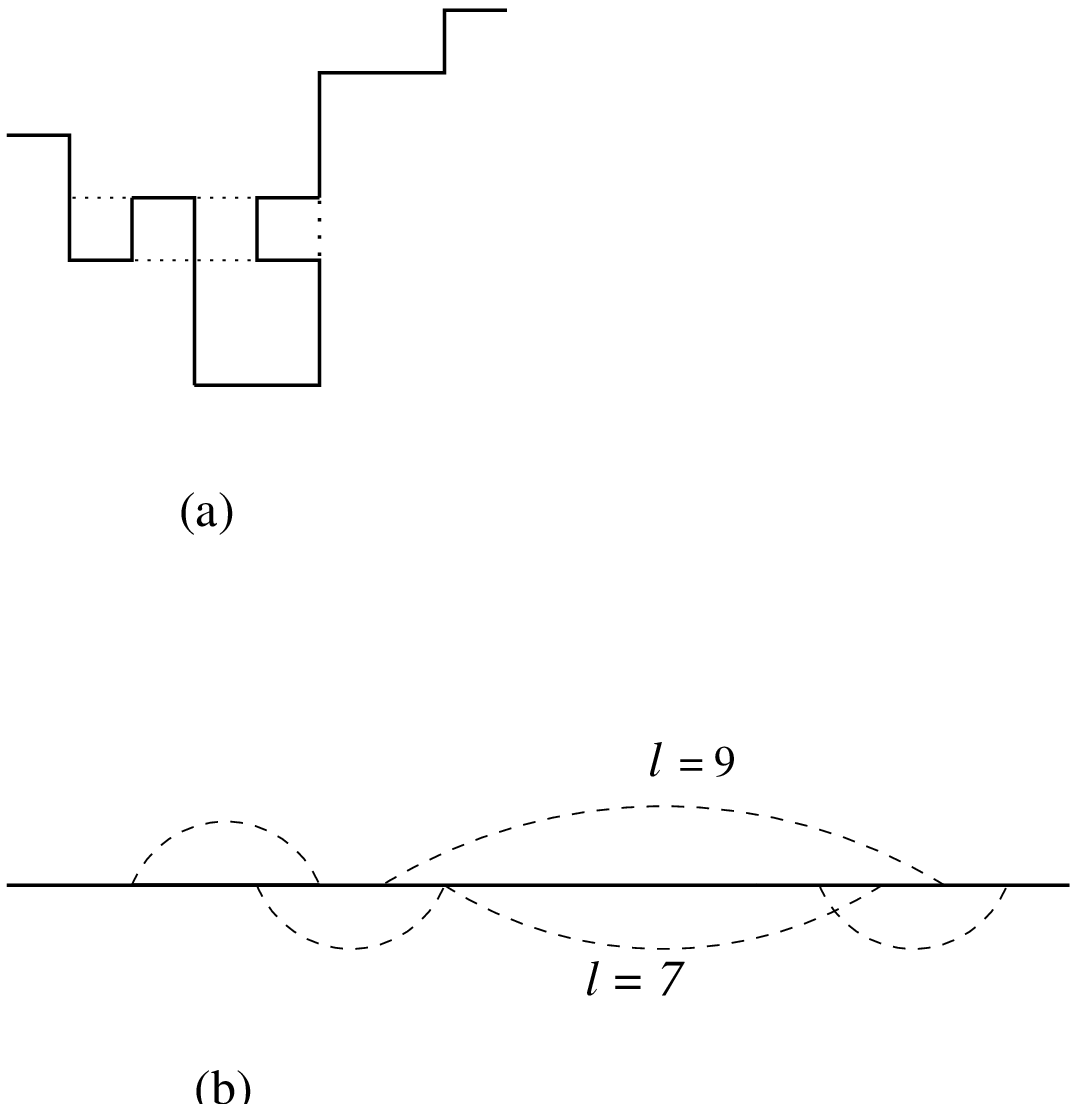}

\psfig{file=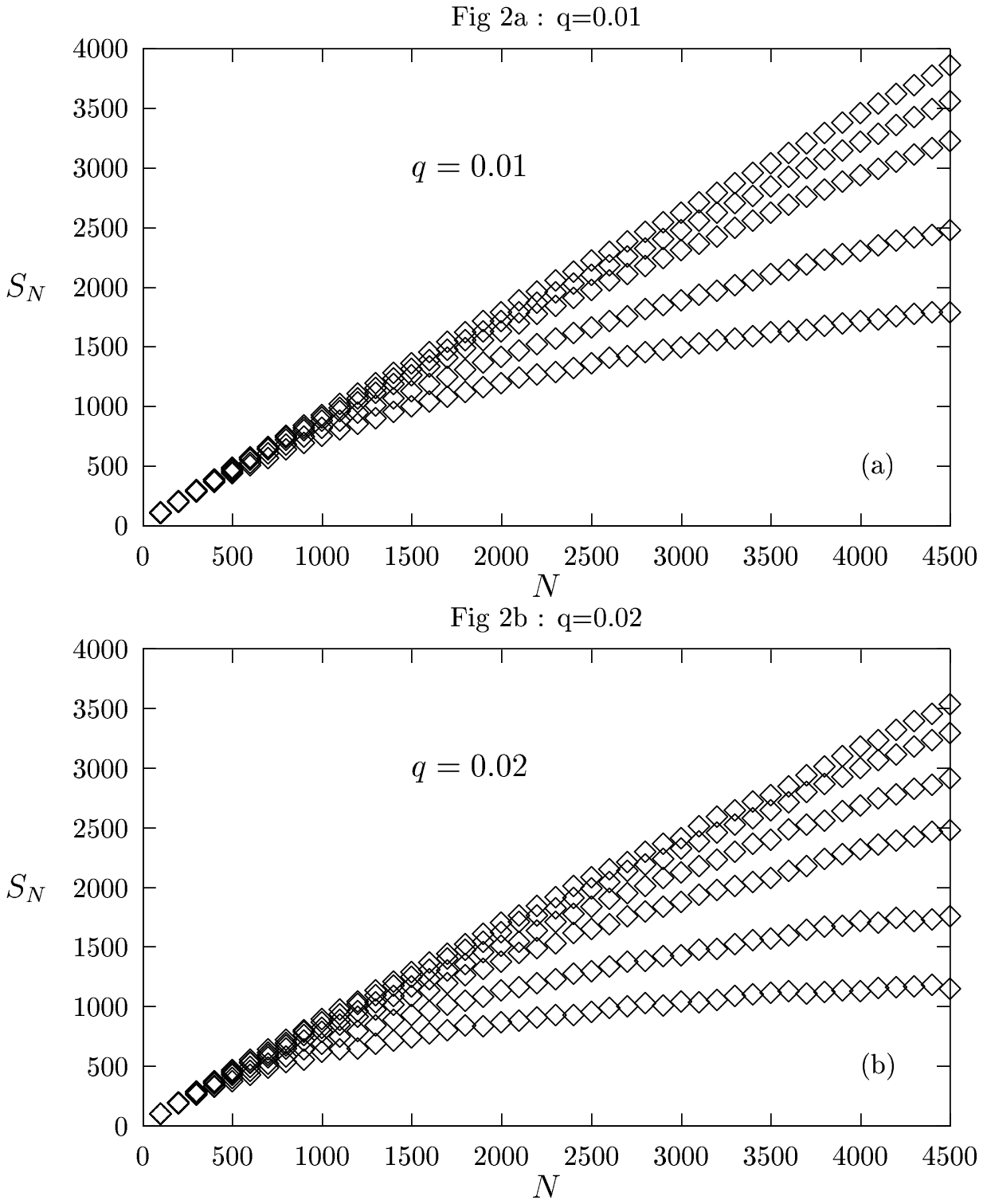}

\psfig{file=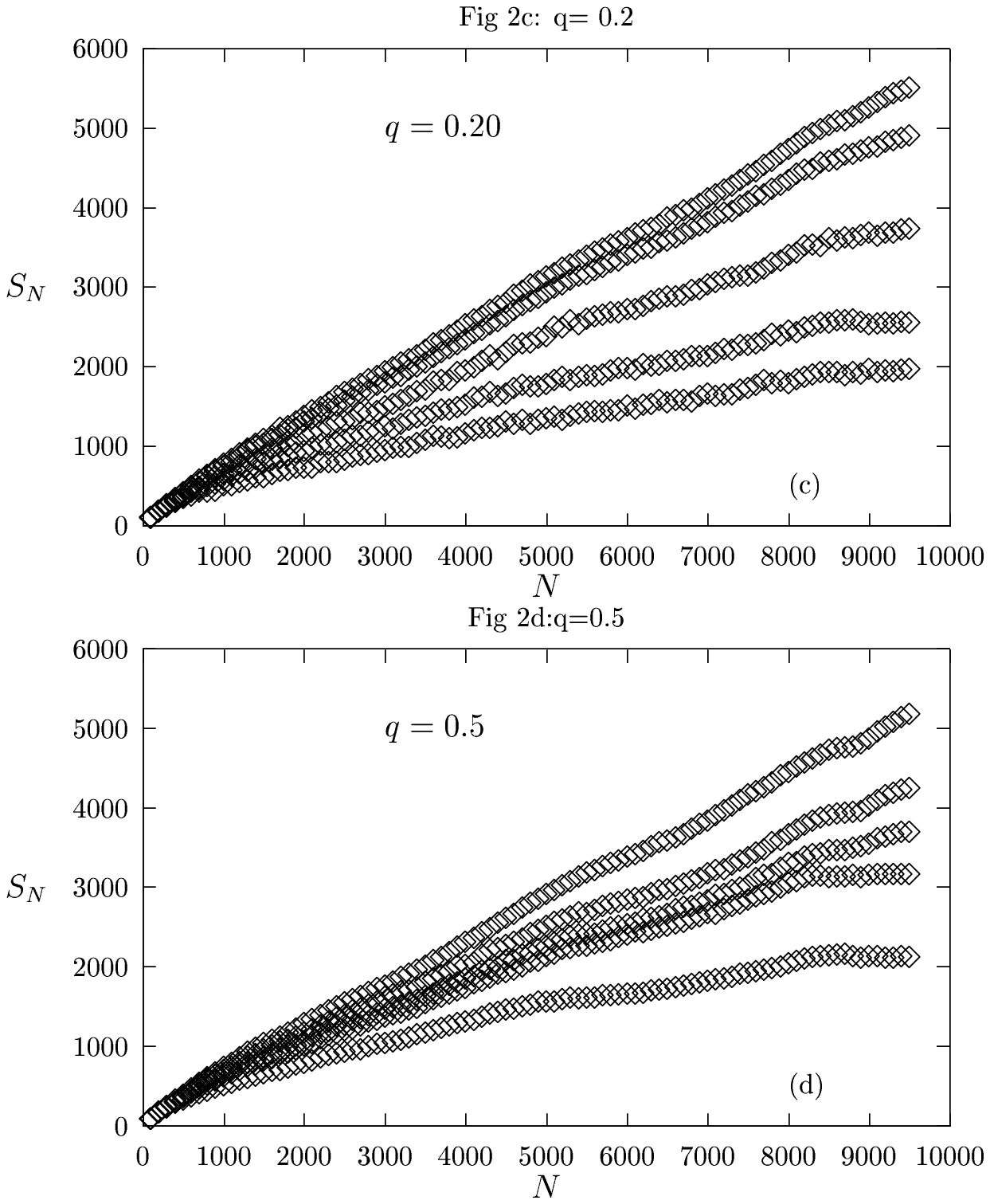}
\psfig{file=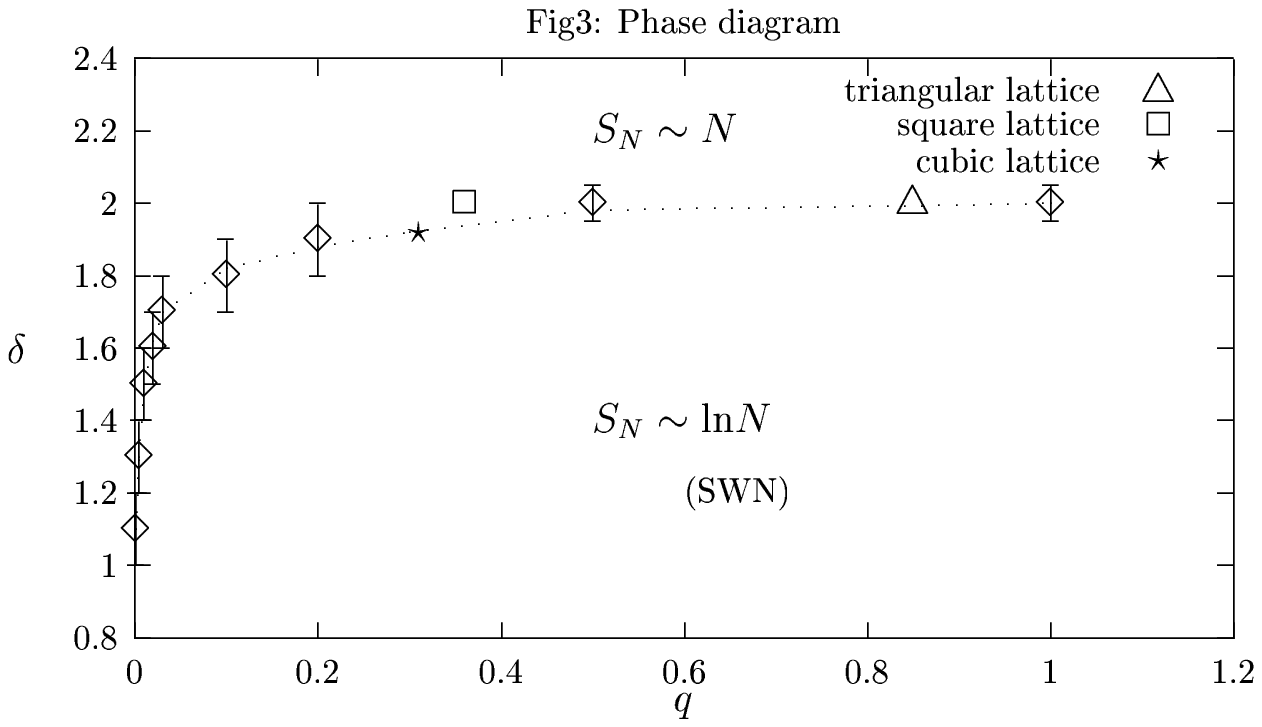}

\end{document}